\documentclass[%
 aip,
 jap,%
 amsmath,amssymb,
reprint,%
]{revtex4-1}

\usepackage{graphicx}
\usepackage{dcolumn}
\usepackage{bm}
\usepackage[mathlines]{lineno}

\usepackage{hyperref}
\usepackage{indentfirst}
\hypersetup{colorlinks=true,linkcolor = blue, citecolor = blue}
\usepackage{eurosym}
\usepackage{caption}
\usepackage{subcaption}
\usepackage{cleveref}
\usepackage{url}

\begin{document}

\preprint{AIP/123-QED}

\title[Sample title]{Extension of the General Thermal Field Equation for nanosized emitters}

\author{A. Kyritsakis}
 \email{akyritsos1@gmail.com.}
\author{J. P. Xanthakis}%
\affiliation{ 
Department of Electrical and Computer Engineering, National
Technical University of Athens, Zografou Campus, Athens 15700, Greece
}

\begin{abstract}

During the previous decade, K.L. Jensen et. al. developed a general analytical model that successfully describes  electron emission from metals both in the field and thermionic regimes, as well as in the transition region. In that development, the standard image corrected triangular potential barrier was used. This barrier model is valid only for planar surfaces and therefore cannot be used in general for modern nanometric emitters. In a recent publication the authors showed that the standard Fowler-Nordheim theory can be generalized for highly curved emitters if a quadratic term is included to the potential model. In this paper we extend this generalization for high temperatures and include both the thermal and intermediate regimes. This is achieved by applying the general method developed by Jensen to the  quadratic barrier model of our previous publication. We obtain results that are in good agreement with fully numerical calculations for radii $R>4nm$, while our calculated current density differs by a factor up to 27 from the one predicted by the Jensen's standard General-Thermal-Field (GTF) equation. Our extended GTF equation has application to modern sharp electron sources, beam simulation models and vacuum breakdown theory. 

\end{abstract}

\maketitle

\section{\label{sec:intro}Introduction}
Most cold field emission (CFE) devices are not actually so "cold" in the sense that despite the absence of external heating the flow of the current itself elevates the device temperature significantly above the room temperature. In general, the amount by which this happens depends on the particular application. In many cases, this heating is enough to reach a condition where we leave the CFE regime and the traditional Fowler-Nordheim (FN) theory \citep{FN1928,Nordheim1928} -even with a temperature correction factor \citep{MurphyG}- is not reliable. On the other hand, this heating is insufficient to reach the thermionic emission regime where the Richardson-Laue-Dushman \citep{Richardson,Dushman} equation is valid. Therefore, an equation describing all three regimes -the CFE, the thermionic emission and intermediate between these two- is necessary. Such a need was recognized by Jensen et. al. \citep{Jensen2006,Jensen2007} who developed an equation appropriate for all three regimes with considerable success. In their work, Jensen et. al. used the image-rounded linear potential barrier or the traditional FN theory, which is commonly known as the Schottky-Nordheim (SN) barrier \citep{Nordheim1928,Schottky1923}. 

\begin{figure}[htbp]
	\centering
    \includegraphics[width=.8\linewidth]{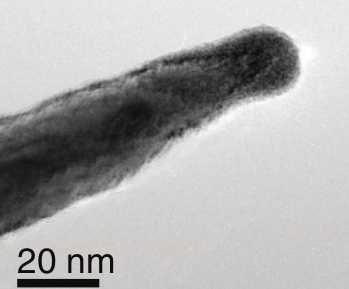}
    \caption{TEM image of a tungsten nano-emitter used in Near Field Emission Scanning Electron Microscopy (reprint from \citep{CabreraPRB}). The radius of curvature at the apex is about 6nm.}
    \label{fig:nanoemitter}
\end{figure}

However during the last ten years the radius of curvature at the apex of the emitting tips has been shrinking and has now reached the values down to 4nm \citep{ZaninAIEP,Guerrera}, which is roughly 2 times the width of the tunneling region. Such a nanometric tip is shown in figure \ref{fig:nanoemitter}. This shrinking apart from increasing the Joule heating effect has another more important effect. The electrostatic potential deviates from the simple linear form and the corresponding transmission coefficient may differ by as much as 2 orders of magnitude \citep{KXnonfn}. It is therefore imperative to revisit the generalized thermal field equation (GTF) -the name adopted by Jensen \citep{Jensen2007}- and extend this equation to systems with emitting tips in the few nanometers range.

Among the many applications where an extended GTF (EGTF) theory is necessary, we wish to mention the vacuum breakdown phenomena, which have always attracted the interest of the scientific community. A significant recent problem in this respect is posed by the construction of a new linear RF accelerator under development in CERN (CLIC) \citep{clic}, whose operation is limited by parasitic vacuum breakdown phenomena. As a result, there is recently an increased interest in the development of reliable simulation tools that explain and predict such phenomena \citep{arcPIC,Eimre2015}. To simulate such phenomena it is important to be able to predict electron emission from metallic protrusions \citep{Eimre2015}, under high electric fields and high temperature in all three regimes. Such protrusions may have radii from hundreds of nm to 4-10nm where the linear approximation to the potential is no longer valid. It is the purpose of this paper to extend the existing GTF theory so as to be applicable to such nanometric sized emitters.

\section{The potential}
\label{sec:pot}
In line with our previous publication \citep{KXnonfn}, we assume an arbitrary, smooth (in the mathematical sense), rotationally symmetric emitting surface with apex radius of curvature $R$. Then the electrostatic potential $\Phi$ along the symmetry axis $z$, when $(z/R)\ll1$,  is approximated by
\begin{equation}
\Phi(z)=Fz\left[1-\frac{z}{R}+O\left(\frac{z}{R}\right)^2\right]
\label{eq:Phi}
\end{equation}
In the above equation, $F$ is the local electrostatic field at the apex ($z=0$) and we have assumed that the emitter is grounded ($\Phi(z=0)=0$). In order to obtain the total barrier potential  energy $U(z)$, we have to add the work function $\phi$ and the image interaction. For the latter we use the standard expression for the image potential for an electron out of a sphere \citep{JensenImage}. We obtain
\begin{equation}
U=\phi-eFz\left(1-\frac{z}{R}+O\left(\frac{z}{R}\right)^2\right)-\frac{Q}{z(1+z/2R)}.
\label{eq:Uofz}
\end{equation}
In the above equation \eqref{eq:Uofz}, $Q=e^2/16\pi\epsilon_0\approx0.36eVnm$ is the standard image pre-factor and $e$ is the elementary charge.

\begin{figure}[htbp]
	\centering
    \includegraphics[width=\linewidth]{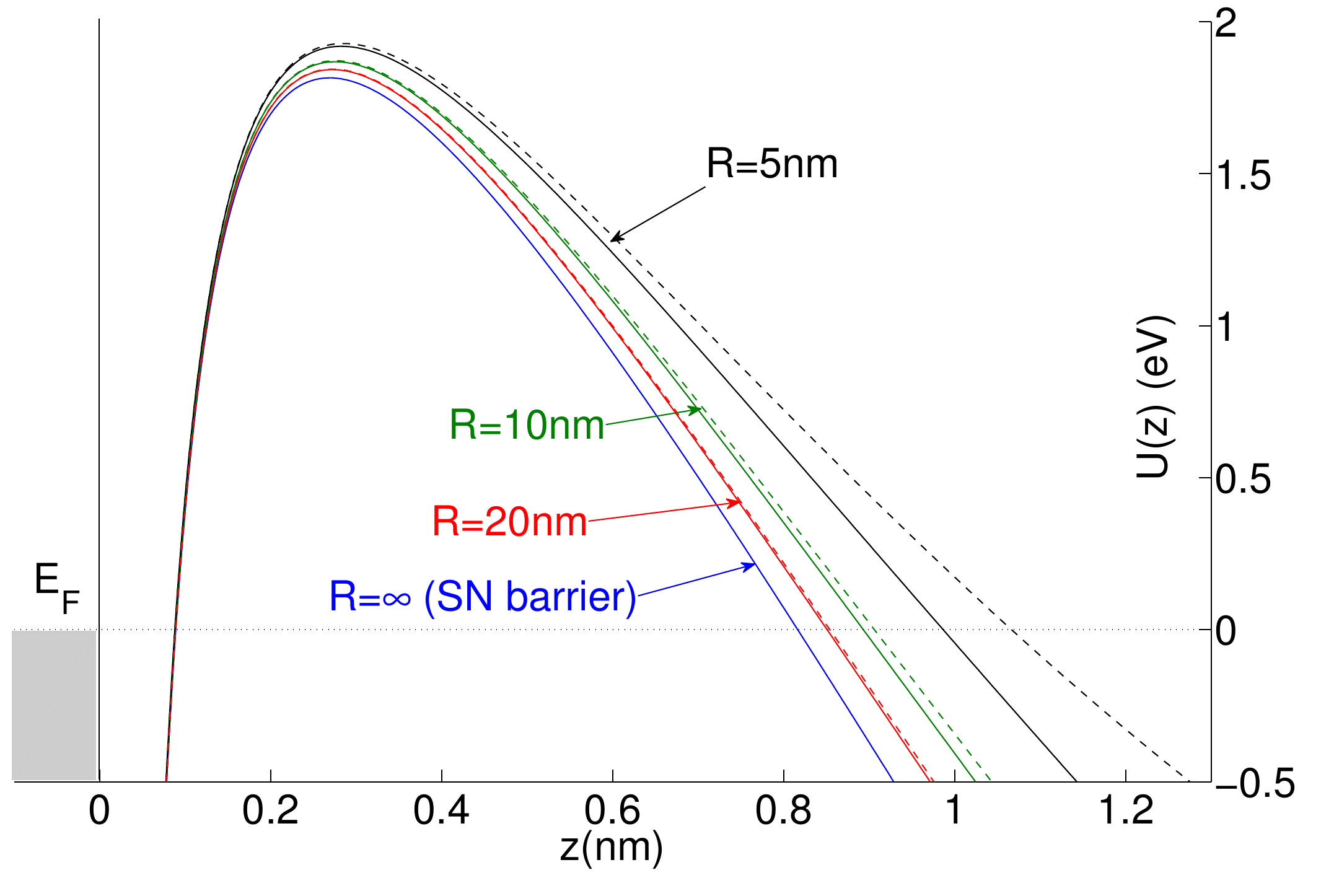}
    \caption{Potential barrier for an ellipsoidal surface as calculated numerically (solid lines) and as calculated by eq.~(\ref{eq:Uofz}) (dashed lines). The apex radius of curvature of such a tip is $R=(R_2)^2/R_1$ where $R_1$ and $R_2$ are correspondingly the major and minor radii of the ellipse. The rest of the parameters are $F=5V/nm,\phi=4.5eV$ and $E_F\equiv 0$.}
    \label{fig:fig1}
\end{figure}

The validity of the approximation of eq.~(\ref{eq:Uofz}) (neglecting $O(z/R)^2$ terms) is shown in fig.~\ref{fig:fig1}. The potential barrier is plotted for $F=5V/nm$, $\phi=4.5eV$ and various $R$, as calculated both numerically by using the spheroidal model of references \citep{KXnonfn,KXBeamspot,KXAPL} (solid lines) and analytically by eq.~(\ref{eq:Uofz}). It can be seen that as $R$ reduces below $20nm$, the potential deviates significantly from the standard SN barrier. However the quadratic approximation of eq.~(\ref{eq:Uofz}) is satisfactory for radii down to about $4-5 nm$.  

In CFE the only energy levels that are of interest are those closely below the Fermi level $E_F$ (throughout this paper $E_F$ will be considered as the zero energy level, i.e. $E_F\equiv 0$). Those energy levels have non-negligible probability to be occupied by electrons with significant tunneling probability. On the contrary, in high temperatures a much wider energy range is of interest. Especially, of significant importance is the maximum point of the barrier $U_{m}$, above which the electrons exit the metal over the barrier and pure thermionic emission occurs. In Jensen's development \citep{Jensen2006,Jensen2007} that maximum point plays a significant role and enters the main equations.

In fig.~\ref{fig:fig1} we can clearly see that the curvature of the emitter significantly changes both the position $z_{m}$ and the value of $U_{m}$ of the maximum, which influences the Gamow exponent. Hence it is essential to obtain an algebraic approximation for $U_{m}$. As in ref. \citep{KXnonfn} we will work with the standard reduced dimensionless variables that are much easier to manipulate: $\zeta=eFz/\phi, x=\phi/eFR, y=2\sqrt{QeF}/\phi$. In terms of those variables eq.~(\ref{eq:Uofz}) becomes
\begin{equation}
\frac{U(\zeta)}{\phi}=1-\zeta\left[1-x\zeta+O\left((\zeta x)^2\right)\right]-\frac{y^2}{4\zeta(1+x\zeta/2)}
\label{eq:Uofzeta}
\end{equation}
In the above equation (\ref{eq:Uofzeta}) $x\ll1$ (see ref.\citep{KXnonfn}) since $\phi/eF$ is a metric of the length of the barrier $L$. If $L/R$ is not small, then there are values of $z/R$ which are not small and more terms in the expansion of eq. \eqref{eq:Phi} have to be taken into account. Provided this is not the case, we may expand given expressions in $x$ and neglect $O(x^2)$ terms. In order to find $\zeta_{m}$, the point were $U_{m}$ occurs, we take the derivative of eq.~(\ref{eq:Uofzeta}) with respect to $\zeta$
\begin{eqnarray}
\displaystyle
&U'(\zeta)=\phi \left( 2 \zeta  x+\frac{y^2}{4 \zeta ^2 \left(\frac{\zeta  x}{2}+1\right)} \right. \nonumber \\ 
&+\left. \frac{x y^2}{8 \zeta  \left(\frac{\zeta  x}{2}+1\right)^2}-1+O((\zeta x)^2) \right).
\label{eq:duzero}
\end{eqnarray}

For the standard SN barrier ($x=0$), the maximum occurs at $\zeta=y/2$. We approximate the root of the above expression by applying the Lagrange inversion theorem\cite{NIST:DLMF} on that point
\begin{equation}
\zeta_{m}=\frac{y}{2}-xy\frac{\phi}{U''(y/2)}+O(x^2).
\label{eq:zmax}
\end{equation}
Note that we write $O(x^2)$ because $\zeta_m$ is now a constant, not a variable. From eq.~(\ref{eq:duzero}) we can obtain $U''(y/2)=-4/y+2x+O(x^2)$ and our final result is
\begin{equation}
\zeta_{m}=\frac{y}{2}+\frac{xy^2}{4}+O(x^2).
\label{eq:zmaxfin}
\end{equation}
\begin{figure}[htbp]
	\centering
    \includegraphics[width=\linewidth]{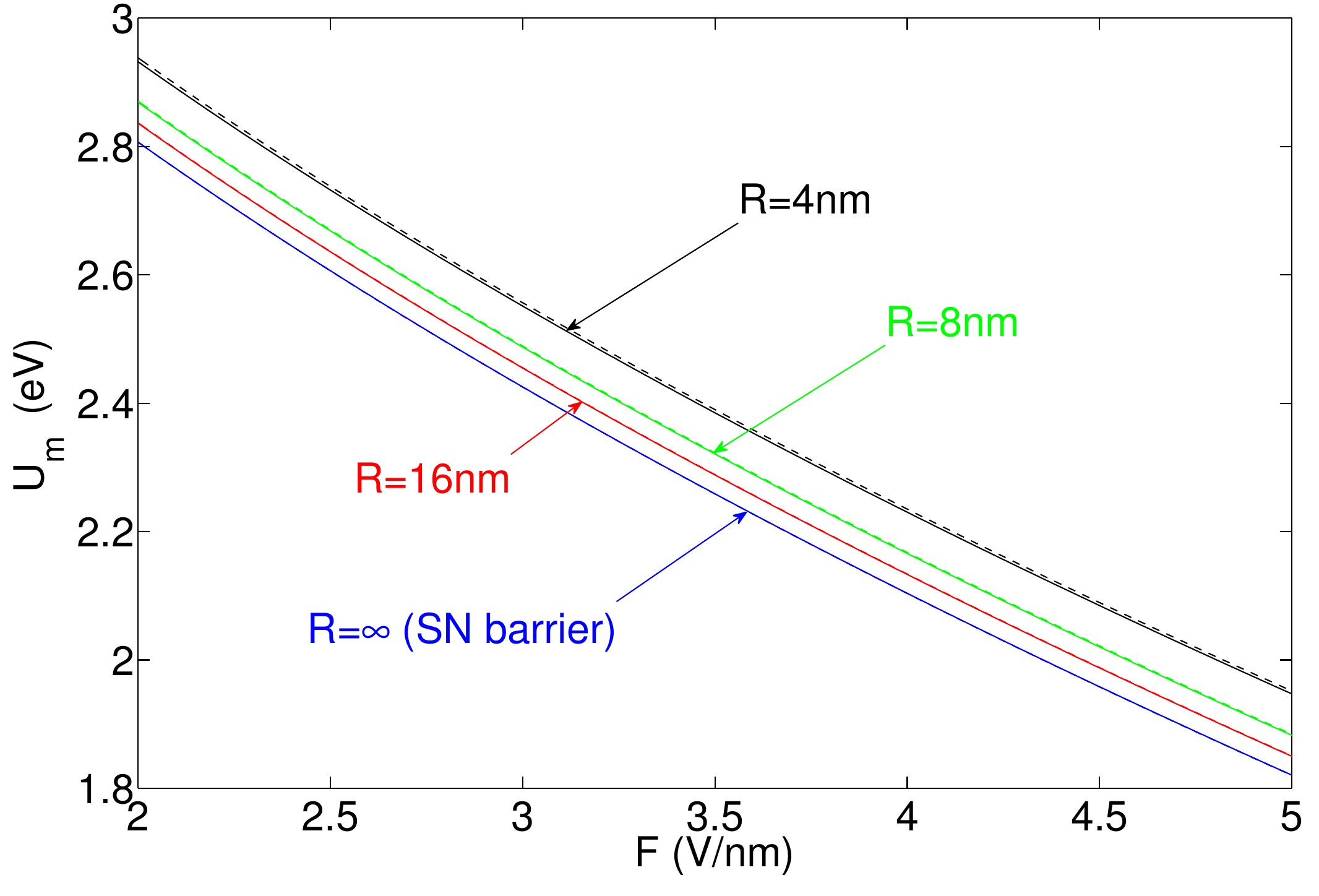}
    \caption{$U_{m}$ as a function of $F$ for various $R$ as calculated numerically by the spheroidal model (solid lines) and by the approximation~(\ref{eq:Umax}). $\phi=4.5eV$.}
    \label{fig:fig2}
\end{figure}

Now by substituting the result of eq.~(\ref{eq:zmaxfin}) into eq.~(\ref{eq:Uofzeta}) we may obtain $U_{m}$
\begin{subequations}
\begin{equation}
U_{m}=\phi\left(1-y+\frac{3 x y^2}{8}+O(x^2)\right) \textrm{, or:}
\end{equation}
\begin{equation}
U_{m}\approx \phi-2\sqrt{eFQ}+\frac{3Q}{4R}
\end{equation}  
\label{eq:Umax}
\end{subequations}
The accuracy of the approximation~(\ref{eq:Umax}) is confirmed numerically in fig.~\ref{fig:fig2}. We may see that the numerically calculated $U_{m}$ differs significantly from the one for the SN barrier. However our approximation is very close to it in the whole range of $F$ and for radii $R$ even below 4nm.

Before closing this section, we should calculate one more important quantity related to the potential which will be used in section \ref{sec:trans}. That is the second derivative of $U$ with respect to $z$ at $z_{m}$. Differentiating eq.~(\ref{eq:duzero}) and substituting $\zeta_{m}$ from \eqref{eq:zmaxfin} we obtain
\begin{subequations}
\begin{eqnarray}
\left.\frac{\partial ^2U}{\partial z^2}\right|_{z_{m}} = \left(\frac{eF}{\phi}\right)^2 U''\left(\zeta_{m}\right)=\nonumber\\
=\frac{(eF)^2}{\phi}\left(-\frac{4}{y}+8x+O(x^2)\right) \textrm{, or}
\end{eqnarray}
\begin{equation}
\left.\frac{\partial ^2U}{\partial z^2}\right|_{z_{m}} \approx -\frac{2(eF)^{3/2}}{\sqrt{Q}}+8\frac{eF}{R} \textrm{ .}
\end{equation}
\label{eq:ddUmax}
\end{subequations}

The various variables and constants used throughout this paper are assembled in table \ref{tab:tab1}.

\begin{table}[htbp]
	\centering
	\caption{Terms, symbols and constants.}
	\label{tab:tab1}	
	\begin{tabular*}{\linewidth}[t]{l @{\extracolsep{\fill}} r}
	\hline \hline
	Symbol & Description \\
	\hline
	
	$Q$ 		& $e^2/16\pi\epsilon_0 \approx 0.35999eVnm$ 				\\[1ex]
	$g$			& $\sqrt{8m}/ \hbar \approx 10.246(nm \sqrt{eV})^{-1}$		\\[1ex]
	$Z_S$		& $em/2\pi^2\hbar^3 \approx 1.618\times10^{-4}A(eVnm)^{-2}$	\\[1ex]
	$x$ 		& $\phi/eFR$ 												\\[1ex]
	$y$ 		& $2 \sqrt{QeF}/\phi$ 										\\[1ex]
	$\zeta$ 	& $eFz/\phi$ 												\\[1ex]
	$\eta$		& $E/U_m$													\\[1ex]
	$\beta$		& $-G'(E_m)$												\\[1ex]
	$B_F$		& $G(0)$													\\[1ex]
	$C_F$ 		& $-G'(0)U_m$ 												\\[1ex]
	$C_q$		& $-G'(U_m)U_m$												\\[1ex]
	$n$			& $1/ (\beta kT)$											\\[1ex]
	$s$			& $G(E_m)+\beta E_m$										\\[1ex]
	$u$			& $G(E_m)-\beta (E-E_m)$									\\[1ex]
	\hline
	\end{tabular*}
\end{table}

\section{\label{sec:trans}The transmission coefficient}
In order to obtain the transmission coefficient we will use the standard Kemble \citep{Kemble} formula in line with Jensen \citep{Jensen2006,Jensen2007} and Murphy and Good \citep{MurphyG}
\begin{equation}
D(E) = \frac{1}{1+\exp(G(E))}
\label{eq:Kemble}
\end{equation}
where
\begin{equation}
G(E)=g \int_{z_1}^{z_2}\!\sqrt{U(z)-E)}\,dz \textrm{,}
\label{eq:Gamow}
\end{equation}
$z_1$ and $z_2$ are the turning points where $U(z)=E$ and $g=2\sqrt{2m}/ \hbar \approx 10.246 (eV)^{-1/2} (nm)^{-1}$. This formula is an improvement to the standard JWKB formula \citep{landau} $D=\exp(-G)$, more appropriate for energies near or above $U_{m}$. The two formulas are asymptotically equal for "deep tunneling" ($G\gg1$) which is dominant in CFE. 

In Jensen's development \citep{Jensen2006,Jensen2007} of the GTF theory, $G(E)$ is approximated for the entire energy range by Hermite polynomial interpolation, after calculation of $G(E)$ and its derivative on the crucial energy levels $E=E_F \equiv 0$ and $E=U_{m}$. The calculations are done for the SN barrier. Here we will approximate those quantities for our generalized barrier. 

For $E=0$ we already have approximations developed in ref. \citep{KXnonfn}
\begin{subequations}
\begin{equation}
G(0)\approx \frac{2}{3}g\frac{\phi^{3/2}}{eF}\left(v(y)+x \omega(y)\right)
\end{equation}
\begin{equation}
G'(0) \approx -g\frac{\sqrt{\phi}}{eF}\left(t(y)+ x \psi(y)\right)
\end{equation}
\label{eq:Gatzero}%
\end{subequations}

where $v(y)$, $t(y)$, $\omega(y)$ and $\psi(y)$ are the correction functions extensively analyzed, tabulated and approximated in ref. \citep{KXnonfn}. The rest of the symbols have their conventional meaning.

For electrons in energy levels near or above $U_{m}$ we will use the parabolic barrier model in line with Jensen. Hence near $z_{m}$ the barrier can be approximated by the parabolic form
\begin{equation}
U=U_{m}-\frac{1}{2}U''(z_{m})z^2
\label{eq:Umaxapp}
\end{equation}
and the Gamow exponent $G$ takes the form \citep{landau}
\begin{equation}
G(E)=-\frac{2\pi\sqrt{m}}{\hbar\sqrt{U''(z_{m})}} (E-U_{m}).
\label{eq:Gnearmax}
\end{equation}
From the above we easily obtain
\begin{equation}
G(U_{m})=0 \textrm{ , } G'(U_{m})=-\frac{2\pi\sqrt{m}}{\hbar\sqrt{U''(z_{m})}}.
\label{eq:Gatmax}
\end{equation}

Now we can easily construct the Hermite interpolation polynomial which interpolates $G(E)$ and its derivative at $0$ and $U_{m}$. In terms of the reduced dimensionless variables $\eta=E/U_{m},B_F=G(0),C_F=-G'(0)U_{m},C_q=-G'(U_{m})U_{m}$ we have (see eq.~(42) of ref. \citep{Jensen2007})
\begin{eqnarray}
&G(\eta)=B_F-C_F\eta \nonumber \\
&+\eta^2\left[(C_F-C_q)(2-\eta)-(B_F-C_q)(3-2\eta)\right].
\label{eq:polyG}
\end{eqnarray}
\begin{figure}[htbp]
	\centering
    \includegraphics[width=\linewidth]{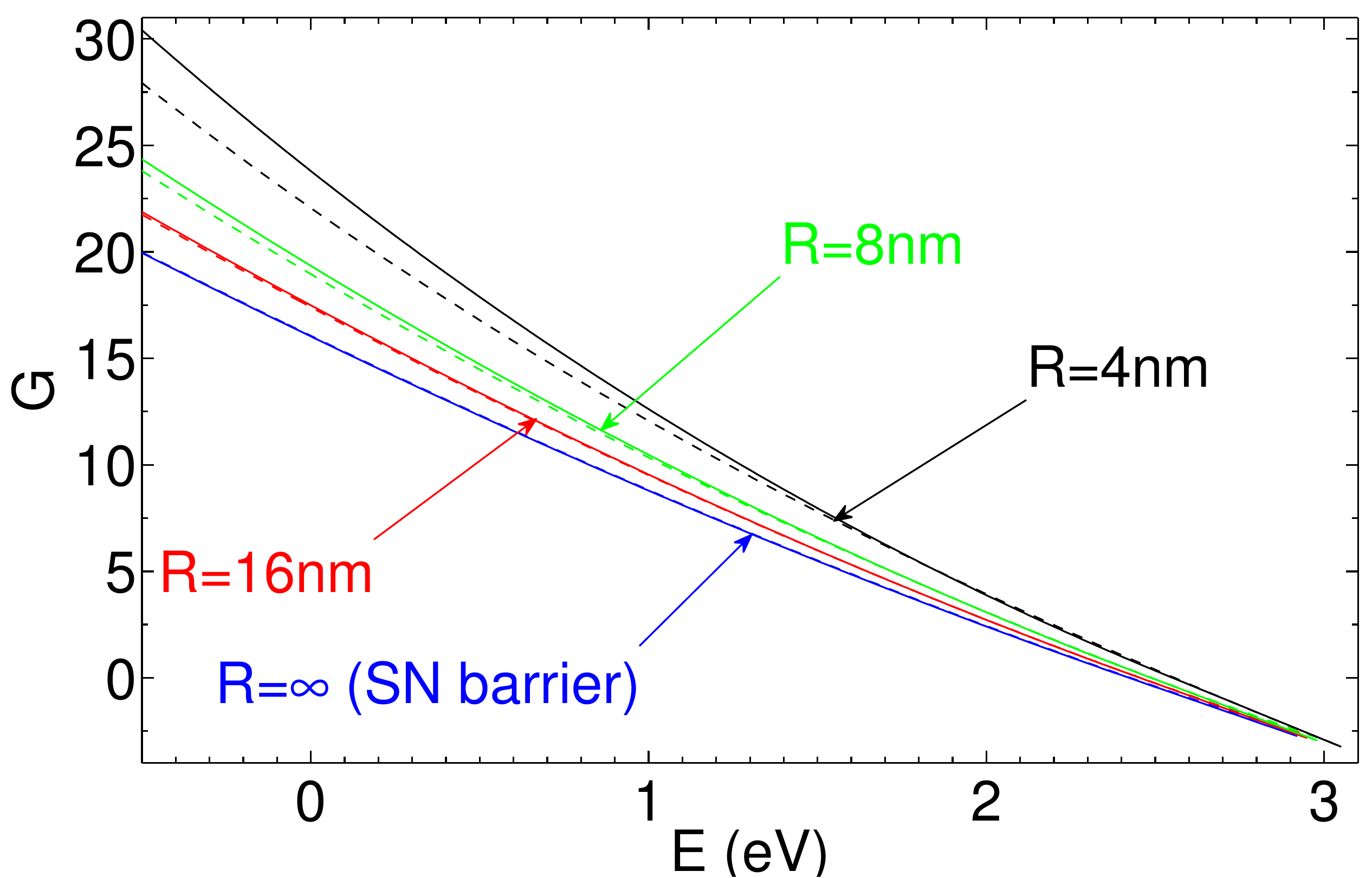}
    \caption{Gamow exponent $G(E)$ for various radii $R$ as calculated numerically (solid lines) and by the algebraic approximation of equations (\ref{eq:Gatzero}-\ref{eq:polyG}). For this figure we have used $F=3V/nm,\phi=4.5eV$.}
    \label{fig:fig3}
\end{figure}

In order to assess the validity of the above approximations, we shall compare $G(E)$ as calculated by equations (\ref{eq:Gatzero}-\ref{eq:polyG}) and as calculated numerically. To calculate $G$ numerically we have calculated $U(z)$ exactly the same way as calculated in figures \ref{fig:fig1} and \ref{fig:fig2} and inserted it into eq.~\eqref{eq:Gamow}. For $E>U_m$ we have used the parabolic model of eq.~(\ref{eq:Gnearmax}). We indeed see this comparison in fig. \ref{fig:fig3}.

We can see that as $R$ reduces below 20nm, $G$ starts to deviate significantly from the one calculated for the SN barrier. Our approximation predicts successfully this deviation, however it starts failing for $R<5nm$. The deviation of the numerically calculated curves from both those for the SN barrier and our approximations, become more pronounced as we go into lower energies. This has been anticipated already from fig.~\ref{fig:fig1}, as we can see that the area between the curves increases for decreasing energy. In general, for energies below the Fermi level, our model is limited by the errors of the approximations of eq.~(\ref{eq:Gatzero}) that are analyzed in ref. \citep{KXnonfn}.

Before closing this section, we assemble the important quantities that are calculated in the previous sections in table \ref{tab:tab2}. We compare the standard expressions for the SN barrier to the generalized ones for our quadratic barrier.

\begin{table*}[htbp]
	\centering
	\caption{Important expressions for the SN and quadratic barriers.}
	\label{tab:tab2}	
	\begin{tabular*}{0.7\textwidth}[t]{l @{\extracolsep{\fill}} c @{\extracolsep{\fill}} r}
	\hline \hline
	Quantity & SN barrier & Quadratic barrier \\
	\hline
	$\Phi(z)$		& $Fz$						& $\displaystyle Fz(1-\frac{z}{R})$		\\[1ex]
	$U(z)$ 		& $\displaystyle \phi-eFz-\frac{Q}{z}$	& $\displaystyle \phi-eFz(1-\frac{z}{R})-\frac{Q}{z(1+z/2R)}$ \\[2ex]
	$U(\zeta)$	& $\displaystyle \phi\left(1-\zeta-\frac{y^2}{4\zeta}\right)$	& $\displaystyle \phi\left[1-\zeta (1-x\zeta)-\frac{y^2}{4\zeta(1+x\zeta/2)}\right]$	\\[2ex]
	$\zeta_m$		& $\displaystyle \frac{y}{2}$				& $\displaystyle \frac{y}{2}+\frac{xy^2}{4}$	\\[2ex]
	$z_m$			& $\displaystyle \sqrt{\frac{Q}{eF}}$		& $\displaystyle \sqrt{\frac{Q}{eF}}+\frac{Q}{eFR}$ \\[2ex]
	$U_m$			& $\displaystyle \phi-2\sqrt{eFQ}$		& $\displaystyle \phi-2\sqrt{eFQ}+\frac{3Q}{4R}$ \\[2ex]
	$U''(z_m)$	& $\displaystyle -2\sqrt{\frac{(eF)^3}{Q}}$	& $\displaystyle -2\sqrt{\frac{(eF)^3}{Q}}+8\frac{eF}{R}$\\[2ex]
	$G(0)$			& $ \displaystyle \frac{2}{3}g\frac{\phi^{3/2}}{eF}v(y)$ &$\displaystyle \frac{2}{3}g\frac{\phi^{3/2}}{eF}\left(v(y)+x \omega(y)\right)$\\[2ex]
	$G'(0)$		& $\displaystyle -g\frac{\sqrt{\phi}}{eF}t(y)$	& $\displaystyle -g\frac{\sqrt{\phi}}{eF}\left(t(y)+x\psi(y)\right)$ \\[2ex]
	\hline
	\end{tabular*}
\end{table*} 

\section{\label{sec:curdens}The current density}
In order to obtain the current density $J$ we have to integrate the transmission probability multiplied with the supply function over the energy range
\begin{equation}
J= Z_S kT \int_{-\infty}^{\infty}~\frac{ln\left(1+\exp\left(-E/kT\right)\right)}{1+\exp \left( G(E) \right)}dE.
\label{eq:Jint}
\end{equation}
In the above eq.~(\ref{eq:Jint}), $k$ is the Boltzmann constant, $T$ is the temperature and $Z_S \equiv em/2\pi^2\hbar^3 \approx 1.618 \times 10^{-4} A(eVnm)^{-2}$ is the Sommerfeld current constant \citep{Sommerfeld}. The lower integral limit is in principle the bottom of the conduction band of the emitter, but for metal emitters the integrand becomes negligible in much higher energy than that. This is due entirely to the fact that the Fermi level is much higher than the bottom of the conduction band. Thus taking $-\infty$ as the lower limit presents no error. However, for semiconducting emitters the Fermi level is close to the edge of the relevant band and the above approximation is not accurate enough. 

The main idea in emission theory is to take a linear approximation of $G(E)$. This linearization is done around $E_F\equiv0$ for CFE \citep{FN1928,MurphyG} and around $U_m$ for thermionic emission \citep{Richardson,Dushman,MurphyG}. Then eq.~(\ref{eq:Jint}) can be integrated algebraically. Nevertheless, Jensen \citep{Jensen2006,Jensen2007} showed that both approximations are insufficient in the intermediate regime. In that case, the maximum of the integrand appears at an energy $E_m$, which migrates from $E_F$ to $U_m$ for decreasing $F$ or increasing $T$. The linear approximation for $G(E)$ has then to be done around $E_m$. In that case eq.~(\ref{eq:Jint}) can be written as \citep{Jensen2007}
\begin{equation}
J=Z_S (kT)^2 n \int_{-\infty}^{\infty}~\frac{ln\left(1+e^{n(u-s)}\right)}{1+e^u}du.
\label{eq:Jint2}
\end{equation}
in terms of the reduced dimensionless variables $n\equiv (\beta kT)^{-1}$ with $\beta \equiv -G'(E_m)$ , $s \equiv G(E_m)+\beta E_m$ and $u=G(E_m)-\beta (E-E_m)$.

In order to approximate $E_m$, Jensen distinguishes three regimes \citep{Jensen2007}. The field regime occurs when $(kT)^{-1}>\beta(E=0)$. Then $E_m$ is very close to $E_F\equiv 0$ and $n>1$. The thermionic regime occurs when $(kT)^{-1}<\beta(E=U_m)$. In that case $E_m$ is approximated by $U_m$ and $n<1$. The intermediate regime is found when $\beta(U_m)\leq (kT)^{-1} \leq \beta(0)$ and $E_m$ can be approximated by the root of the equation $\beta(E_m)=1/kT$, i.e. by the point where $n=1$. That point can be found by solving the second order equation
\begin{eqnarray}
&C_q\eta+C_F(1-\eta)+\nonumber \\
&3(2B_F-C_q-C_FN)\eta(1-\eta)=U_m/kT
\label{eq:findEm}
\end{eqnarray}
and choosing the root $\eta_m$ which is between $0$ and $1$. Then of course $E_m$ is found easily by $E_m=U_m\eta_m$.

The integral of eq.~(\ref{eq:Jint2}) is algebraically solvable in terms of series expansions. The result can be approximated depending on the regime by \citep{Jensen2007,Jensen2008}
\begin{equation}
J\approx \left\{
\begin{array}{l r}
J_F/n^2+J_T\quad & (n<1)\\
Z_S (kT)^2 (s+1)e^{-s} \quad & (n=1) \\
J_F+n^2J_T\quad & (n>1)
\end{array}\right.
\label{eq:Jbranches}
\end{equation}
where $J_F$ and $J_T$ are the field and thermal components of the total current that dominate in each regime correspondingly. They are given by
\begin{subequations}
\begin{equation}
J_T = Z_S (kT)^2\Sigma(n)e^{-ns},
\end{equation}
\begin{equation}
J_F = Z_S \left(\frac{1}{\beta^2}\right)\Sigma\left(\frac{1}{n}\right) e^{-s}.
\end{equation}
\label{eq:Js}%
\end{subequations}
$\Sigma(x)$ is a function that can be expressed analytically in terms of the infinite series
\begin{equation}
\Sigma (x)=1+2x^2 \sum\limits_{j=1}^{\infty}~\frac{(-1)^{j+1}}{j^2-x^2}.
\label{eq:Sigma}
\end{equation}
Note that the above definition of $\Sigma(x)$ is derived directly from equations (35) and (31) of ref. \citep{Jensen2007}. This representation is simpler and more general than eq.~(38) of ref. \citep{Jensen2007} as in the latter the series does not converge for $x\geq1$. However it is much easier and accurate enough to use rational function approximations obtained by Jensen et. al. \citep{Jensen2008}
\begin{equation}
\Sigma(x) \approx \frac{1+x^2}{1-x^2}-0.039x^2 \left(9.1043+2.7163x^2+x^4 \right).
\label{eq:sigmapoly}
\end{equation}

\begin{figure}[htbp]
	\centering
    \includegraphics[width=\linewidth]{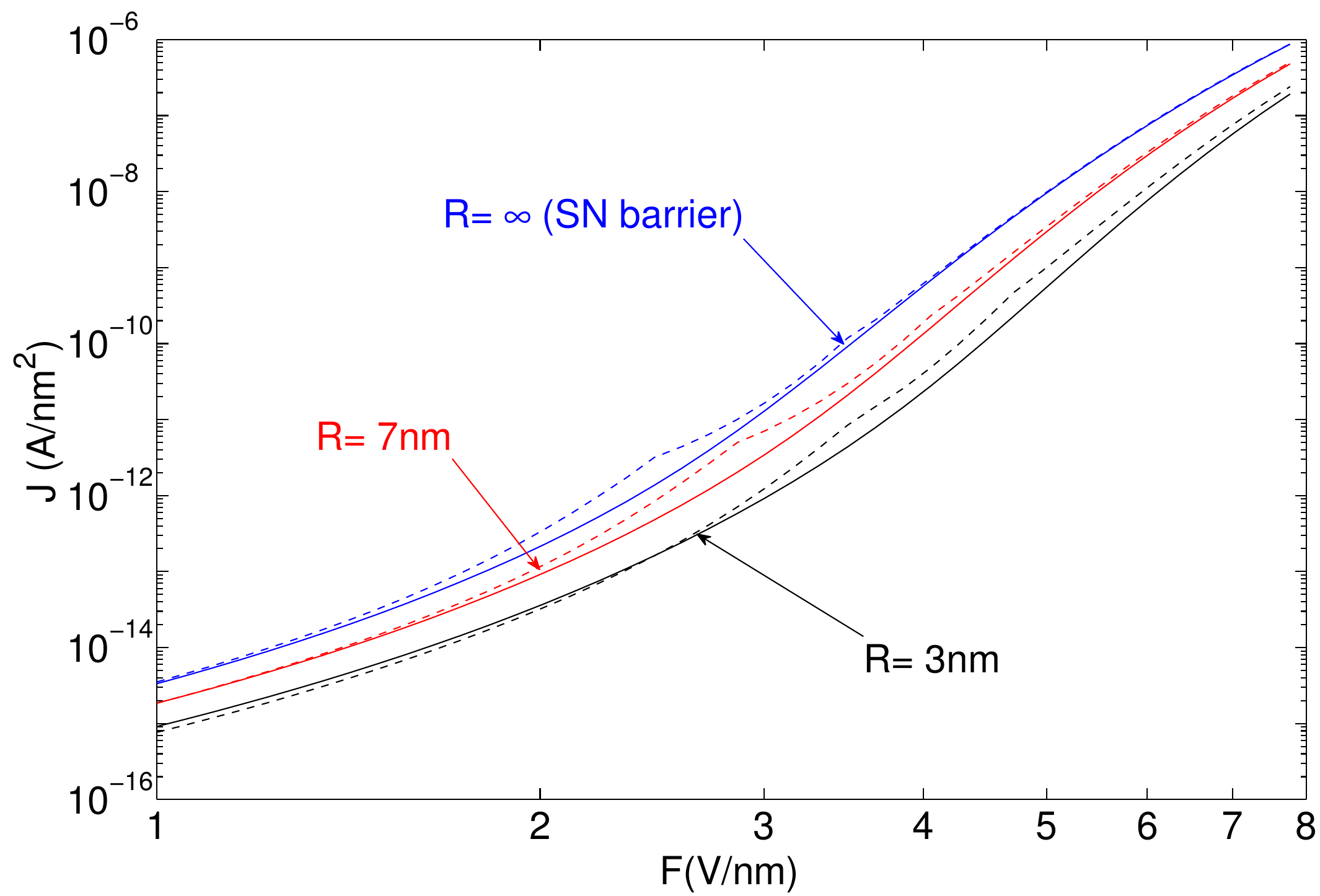}
    \caption{Current density $J$ as a function of the tip field $F$ for various radii $R$. Again solid lines correspond to fully numerical calculation and dashed lines to eq.~\eqref{eq:Jbranches}. Both axes are in logarithmic scale. The work function is $\phi=4.5eV$ and the temperature is $T=1800K$.}
    \label{fig:fig4}
\end{figure}

By implementing the method described above, we may obtain the current density $J$ predicted by the EGTF equation as a function of the local field $F$, work function $\phi$, temperature $T$ and local radius of curvature $R$. In fig.~\ref{fig:fig4} we plot the current density as obtained by eq.~\eqref{eq:Jbranches} (dashed lines) for various radii. We also give the fully numerical calculations (solid lines) for the purposes of comparison. The latter are performed by numerically integrating eq.~\eqref{eq:Jint}, after obtaining $G(E)$ the same way as in fig.~\ref{fig:fig3}.

\section{Discussion}
\label{sec:disc}

In fig.~\ref{fig:fig4} we can see again that as $R$ reduces below 20nm the current density is significantly lower than the one obtained for the SN barrier by Jensen's GTF theory, even in the thermionic emission regime. This difference is more pronounced in the intermediate regime where the numerically calculated $J$ of the $R=3nm$ curve of fig.~\ref{fig:fig4} is up to 27 times lower than the corresponding algebraic one for the SN barrier. However, our extension for the quadratic barrier seems to successfully predict this reduction for radii down to $3nm$. As we go to smaller radii, our theory starts deviating significantly from the numerically calculated values, exhibiting a maximum deviation of about a factor of 2.3 in the intermediate regime for $R=3nm$. This is still much better than the deviation exhibited by the original GTF theory. 

Here a comment is worthy on the apparent improved performance of EGTF in comparison to the results of ref.~\citep{KXnonfn} for pure CFE. From fig.~3 of ref.~\citep{KXnonfn} we see that our approximation for $J$ deviates significantly from the corresponding numerical calculation even for the higher radius of $5nm$. For example, a difference of a factor of about 3 is found for $F=3V/nm$. In contrast, in the general thermal-field regime of fig.~\ref{fig:fig4} our approximation can go down to the radius of $R=3nm$ without deviating from the corresponding numerical calculation for more than a factor of 2.3. 

This finding seems surprising \textit{prima facie}{, but it can be explained by a closer examination of fig. \ref{fig:fig1}. We can see that the approximate curves deviate more from the numerical ones as we go to lower energies. Given the fact that the transmission coefficient $D$ depends on the area under the barrier and above the electron's energy level, it is evident that $D$ is better approximated for higher energy electrons. As a result, our approximation for the transmission coefficient gets better for higher temperatures when electrons occupy higher energy levels. 
\begin{figure}[htbp]
	\centering
    \includegraphics[width=\linewidth]{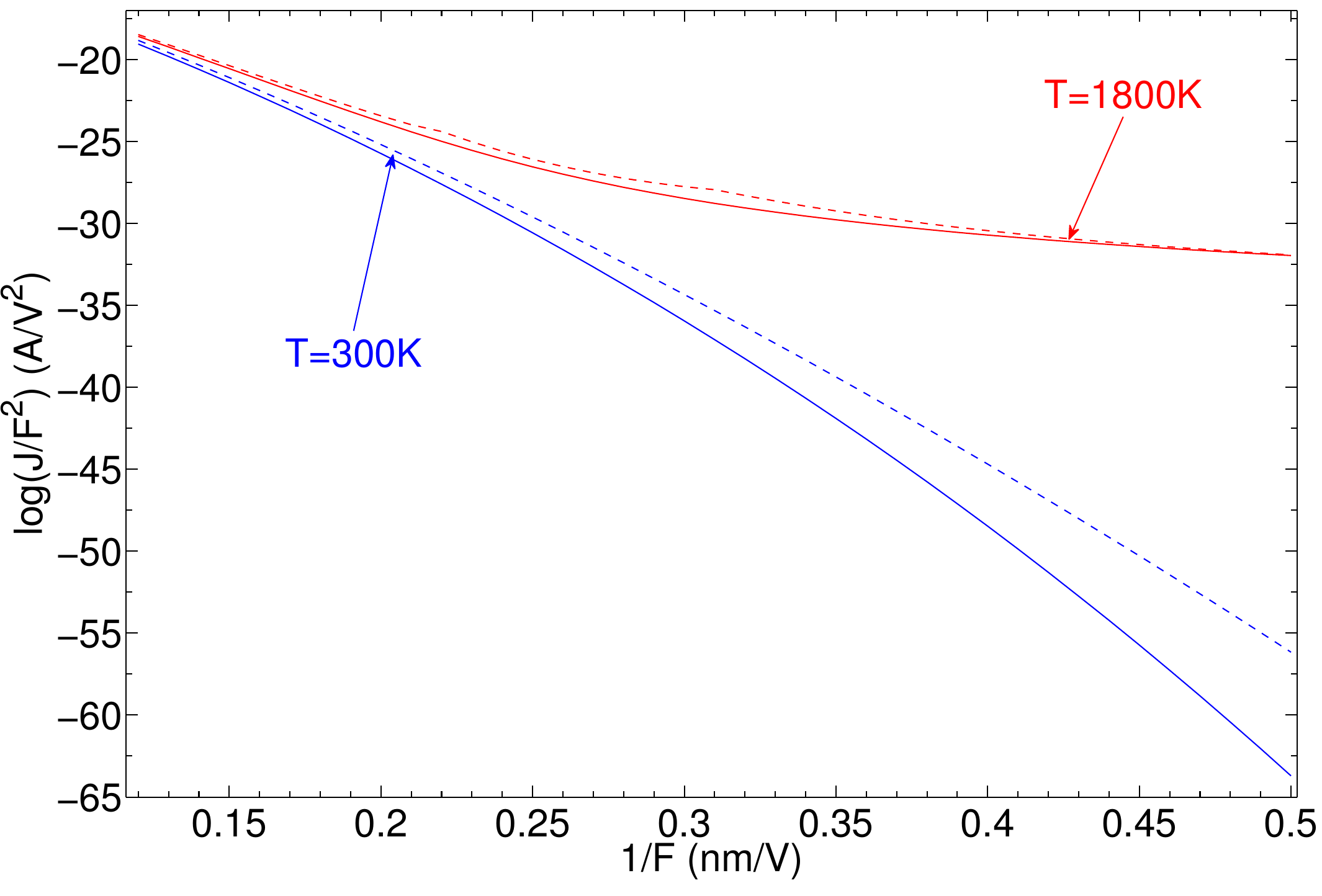}
    \caption{Current density $J$ as a function of the tip field $F$ in the form of F-N plots for two different temperature regimes. Again solid lines correspond to numerical calculation and dashed lines to to eq.~\eqref{eq:Jbranches}. The work function is $\phi=4.5eV$ and the radius $R=4nm$.}
    \label{fig:fig5}
\end{figure}

A better understanding of this behavior is given in fig. \ref{fig:fig5}, where we compare the performance of our approximations for pure FE conditions (blue line , $T=300K$) and for GTF conditions (red line, $T=1800K$) for a small radius of $R=4nm$. For cold FE the approximation tends to deteriorate as we go to lower fields when wider parts of the barrier become of importance. This is evident in the blue curves that turn downwards and start significantly deviating from each other as we go to low fields. However, if the temperature is high, reducing $F$ means that we enter the GTF regime and the majority of the electrons come from near or above the top of the barrier where we still have a valid approximation. Thus the F-N plot turns upwards, and the approximation starts improving as we approach the pure thermal regime. In that regime, the only barrier parameter that is of importance is the height of the barrier $U_m$ and that is very well approximated by eq.~\eqref{eq:Umax} as we can see from fig.~\ref{fig:fig2}.

Finally we wish to make a comment on the accuracy of the calculations of section \ref{sec:curdens} for the current density. We note that the approximate curves of fig. \ref{fig:fig4} exhibit a peak at the transition point between thermionic and intermediate regimes. Around that peak, they significantly deviate from the numerically calculated curves (by about a factor of 2 at most). This error is attributed mainly to errors in the approximation of the integral \eqref{eq:Jint}, regardless of the barrier model that is used. By numerically checking the successive approximations described in the previous section, we found that the error comes mainly from the approximation of finding $E_m$. In fact, if we obtain $E_m$ (and hence $n,s$) numerically, we may obtain a much smaller error in the calculation of $J$

\section{\label{sec:conc}Conclusions}
In conclusion, we have obtained an analytical extension of the general thermal field emission equation appropriate for nanoscopic emitters with radii of curvature down to 3-4nm. We have found that when the emitter has a curvature of $R=3nm$, the emitted current density can be up to 27 times less than the one predicted by the standard GTF equation. Our extension can predict this with an error up to a factor of 2.3. We also find that our approximation is better for high temperatures, and in the GTF regime has a better performance than in the CFE regime.  

\section*{References}

\bibliography{../../bibliography/abbreviated}
\bibliographystyle{aipnum4-1}

\end{document}